\documentclass[abbrv,aps,prb,twocolumn,showpacs,preprintnumbers,amsmath,amssymb,
groupaddress,nofootinbib]{revtex4}

\usepackage{amsmath,amssymb,amsfonts}
\usepackage{bm,graphicx,color}

\newcommand{\gaassb}{$\mathrm{GaAs_{1-y}Sb_y }$}

\newcommand{\cc}[1]{ {\it #1} }
\newcommand{\del}[1]{} 

\renewcommand{\vec}[1]{\mathbf{#1}}

\begin{document}
\title{Excitonic fine structure splitting in type-II quantum dots}
\author{V.~K\v{r}\'apek}
\email{vlastimil.krapek@ceitec.vutbr.cz}
\affiliation{Central European Institute of Technology, Brno University of Technology, Technick\'a 10, 616 00 Brno, Czech Republic}
\author{P.~Klenovsk\'y}
\affiliation{Institute of Condensed Matter Physics, Masaryk University,
Kotl\'a\v{r}sk\'a 2, 611 37 Brno, Czech Republic}
\affiliation{Central European Institute of Technology, Masaryk University, Kamenice 753,
625 00 Brno, Czech Republic}
\author{T.~\v{S}ikola}
\affiliation{Central European Institute of Technology, Brno University of Technology, Technick\'a 10, 616 00 Brno, Czech Republic}
\affiliation{Institute of Physical Engineering, Brno University of Technology, Technick\'a 2, 616 69 Brno, Czech Republic}
\date{\today}

\begin{abstract}
Excitonic fine structure splitting in quantum dots is closely related
to the lateral shape of the wave functions. We have studied theoretically
the fine structure splitting in InAs quantum dots with a type-II confinement
imposed by a GaAsSb capping layer. We show that very small values
of the fine structure splitting comparable with the natural linewidth
of the excitonic transitions are achievable for realistic quantum
dot morphologies despite the structural elongation and the piezoelectric
field. For example, varying the capping layer thickness allows for a fine
tuning of the splitting energy.  The effect is explained
by a strong sensitivity of the hole wave function to the morphology of the
structure and a mutual compensation of the electron and hole anisotropies.
The oscillator strength of the excitonic transitions in the studied quantum
dots is comparable to those with a type-I confinement which makes the 
dots attractive for quantum communication technology as
emitters of polarization-entangled photon pairs.
\end{abstract}
\maketitle

\section{Introduction}

Excitonic fine structure splitting (FSS) refers to a tiny energy splitting
of two bright exciton states confined in quantum dot
(QD) heterostructures with a typical magnitude ranging from units to hundreds
$\mu$eV. It is manifested in a doublet structure
of the exciton recombination band.
It was observed for the first time in GaAs/AlGaAs quantum wells
with fluctuating thickness~\cite{PhysRevLett.76.3005} and then in various QD
systems.~\cite{PhysRevB.65.195315,ISI:000083233000010,PhysRevLett.82.1780,
PhysRevB.81.121309} Soon after its discovery it was attributed to the
electron-hole exchange interaction~\cite{PhysRevB.62.16840}
and its finite value was related to the reduced symmetry, which
needs to be lower than $D_{2d}$.~\cite{PhysRevB.65.195315}

The interest in FSS is triggered by both fundamental and application point
of view. FSS helps to distinguish the spectral features originating
in the recombination of exciton (doublet), biexciton
(doublet with opposite polarization-energy dependence), and trion (singlet).
Benson's proposal of the source of entangled photon pairs relying on zero
FSS~\cite{PhysRevLett.84.2513} has called for the preparation of QD systems with low FSS.
Using (111) substrates for the growth of InAs QDs reduced both structural asymmetry
and piezoelectric contribution.~\cite{PhysRevB.80.161307}
Another attempt involved strain-free
GaAs/AlGaAs QDs with zero piezoelectric field, which, however, still
exhibited a finite FSS due to structural elongation.~\cite{PhysRevB.78.125321,PhysRevB.81.121309}
Post-growth annealing \cite{PhysRevB.69.161301} of InAs QDs allowed to
decrease FSS from $96\ \mathrm{\mu eV}$ to mere $6\ \mathrm{\mu eV}$.
Another class of approaches is based on in-operation tuning, where
the originally large value of FSS is reduced by applying the external
field: electric,~\cite{:/content/aip/journal/apl/90/4/10.1063/1.2431758,:/content/aip/journal/apl/91/5/10.1063/1.2761522}
magnetic,~\cite{PhysRevB.65.195315,PhysRevB.73.033306} or strain. The external strain field allowed to reach
FSS below experimental resolution in GaAs/AlGaAs QDs;~\cite{PhysRevB.83.121302,ISI:000302003600007}
the simultaneous application of electric field allowed for a more powerful symmetry
restoration and rather universal recovery of low FSS.~\cite{PhysRevLett.109.147401}

Various effects contributing to the FSS can be divided into two classes
based on the involved length scale: atomic and macroscopic.
Atomic-scale effects are connected with the irregularities of the
crystal lattice such as the interfaces, particular elements distribution in alloys,~\cite{PhysRevB.86.161302}
charged defects,~\cite{PhysRevB.85.205405} etc.
The magnitude of these effects is still
subject of investigation; the atomistic simulations based on the tight-binding
method~\cite{PhysRevB.88.155319} predict considerably larger values than
those relying on the empirical pseudopotential method.~\cite{PhysRevB.84.241402}
In general, atomic-scale effects are weak compared to those on macroscale.
For example, the magnitudes of about 1~$\mu$eV are reported for a specific alloy
distribution in the AlGaAs barrier~\cite{PhysRevB.86.161302} of GaAs QDs.
The effect is more pronounced when the dot material is an alloy, which should
be therefore avoided when aiming at low FSS. A lower bound of several~$\mu$eV was predicted for 
strain-tuned FSS in ternary $\mathrm{In_{0.6}Ga_{0.4}As}$ QDs.~\cite{PhysRevLett.104.196803}
By macroscopic scale we mean for the purpose of the foregoing discussion
that the characteristic length of the effect is comparable with the dimensions of a QD and
the underlying crystal lattice is perceived as a homogeneous environment.
Thus, the crystal symmetry is no longer relevant and the finite values of FSS
are now related to the symmetry lower than $C_4$, i.e., to the lateral elongation
of the wave functions (e.g. envelope functions of the $\mathrm{k \cdot p}$ theory).
Principal contributions to the FSS on macroscopic scale
arise from the asymmetric (elongated) shape of a QD and piezoelectric field.
Further, it is possible to use the external strain field to induce the anisotropic
effective mass tensor and modify the elongation of the hole wave functions
and the related value of FSS.~\cite{PhysRevB.83.121302}

Further information is contained in the polarization properties of the
exciton doublet. Simple considerations assuming a purely heavy-hole
exciton in an elliptic-disk-shaped QD~\cite{PhysRevB.62.16840} predicted a linear polarization
of both transitions with the low-energy
component polarized parallel with the long QD axis and the
high-energy component having the orthogonal polarization.
Typically, both structural-elongation
and piezoelectric axes are parallel with the crystal axes [110] and
[1$\bar{1}$0], and so
are the polarizations of both components. However, in some structures
with shallow irregular confinement potential, such as quantum well
thickness fluctuations, stochastic polarization directions were
observed.~\cite{PhysRevB.81.121309} Further, when the light-hole contribution to the exciton
ground state becomes important, the polarization orthogonality of
both components is lost.~\cite{PhysRevB.83.121302,PhysRevB.87.075311}

We focus here on QDs with type-II confinement, in which one type
of charge carriers is confined in QD volume and the other in the barrier
close to the QD vicinity. The particular system of interest are InAs QDs
with a thin $\mathrm{GaAs_{1-y}Sb_y}$ overlayer embedded in GaAs.
One reason for selecting this material
system is the possibility to induce a smooth crossover between type-I and
type-II confinement regime simply by changing $y$; the crossover values
between 0.14 and 0.18 have been reported.~\cite{:/content/aip/journal/jap/99/4/10.1063/1.2173188,ISI:000284545200055,1742-6596-245-1-012086}
The other is that it belongs to the minority of systems with
\textit{holes} bound outside. Owing to their large effective mass the holes
are more susceptible to the local potential profile or external perturbations,
offering a larger potential for tuning their wave functions and the related
FSS. The photoluminescence of \gaassb{} capped QDs is rather intense
despite the type-II confinement with the radiative lifetimes as low as 10~ns.~\cite{:/content/aip/journal/apl/101/25/10.1063/1.4773008}
The strain-reducing effect of \gaassb{} layer together with the surfacting effect of antimony
allow to increase the emission wavelengths of standard InAs QDs and 
reach the telecommunication wavelength of 1.3 and 1.55~$\mathrm{\mu m}$.~\cite{:/content/aip/journal/apl/87/20/10.1063/1.2130529,Hospodkova20101383}
Various shapes of \gaassb{} QDs have been reported, including a lens~\cite{:/content/aip/journal/apl/99/7/10.1063/1.3624464}
or a pyramid with a graded In concentration.~\cite{:/content/aip/journal/apl/90/21/10.1063/1.2741608}
Notably, the hole wave function is expected to be composed of two segments
localized in the minima of the piezoelectric potential.~\cite{ISI:000284545200055}

In this work we present a theoretical study of excitonic fine structure
splitting of InAs QDs with $\mathrm{GaAs_{1-y}Sb_y}$ overlayer.
We propose a method to tune the FSS by setting the thickness of the
$\mathrm{GaAs_{1-y}Sb_y}$ layer. The values comparable 
with the natural linewidth can be achieved even in low-symmetry QDs.
The paper is organized as follows: In Section~\ref{s2} a theory of FSS is
described. To gain a qualitative understanding
of the relations between the wave functions and FSS we discuss in Section~\ref{s3} a simplified single band model with Gaussian wave functions.
The full calculations are presented in Section~\ref{s4}. We summarize and conclude
in Section~\ref{s5}.

\section{Theory}
\label{s2}
The single particle states were calculated within the eight-band $\mathrm{k\cdot p}$
theory,~\cite{bastard1981,stier1999} in which the wave functions are expanded into products
of periodic parts of Bloch functions $u_b$ in the $\Gamma$ point and corresponding envelope
functions $\chi_b$,
\begin{equation}
\psi(\vec{r})=\sum_{b\in\{x,y,z,s\}\otimes\{\uparrow,\downarrow\}}
u_\mathit{b} (\vec{r}) \chi_\mathit{b} (\vec{r}).
\end{equation}
In this equation $b$ is the band index, the bands $x$,
$y$, $z$ correspond to the valence band Bloch waves which are antisymmetric
with respect to the corresponding mirror plane and $s$ corresponds
to the conduction band Bloch wave. Following usual conventions, $z$
denotes the growth direction.
The calculations include the effects of the elastic strain via the Pikus-Bir
Hamiltonian~\cite{pikus1960} and the piezoelectric field.
The numeric simulations were performed with Nextnano 3D~\cite{4294186} which employs
the finite difference method.
The simulation space was discretized
with a step of 1 nm.

Once the single particle states are calculated, it is convenient to use them
as a basis for the exciton state $|X\rangle$. First the Slater determinants
$|X(\mathit{ci}, \mathit{vj})\rangle=c_{ci}^\dagger c_\mathit{vj}|0\rangle$ are formed,
where $|0\rangle$ is Fermi vacuum state (empty quantum dot),
$c_{ci}^\dagger$ creates an electron in $i$th conduction state
and $c_{vj}$ annihilates an electron in $j$th valence state, the corresponding single-particle
wave functions are denoted $\psi_\mathit{ci}$, $\psi_\mathit{vj}$, respectively.
For the calculations of FSS we used four Slater determinants formed
from the ground hole and electron states.
Following Ref.~\onlinecite{PhysRevB.62.16840} (Eq.~2.3) the Hamiltonian matrix elements
read
\begin{multline*}
\langle X(\mathit{ci}, \mathit{vj})| \hat{H} |X(\mathit{ck}, \mathit{vl})\rangle = \\
(E_i-E_j) \delta_\mathit{ik} \delta_\mathit{jl}+C(\mathit{ci},\mathit{vj},\mathit{ck},\mathit{vl})+\mathit{EX}(\mathit{ci},\mathit{vj},\mathit{ck},\mathit{vl})
\end{multline*}
where $E_i$ is the energy of $\mathit{i}th$ single particle state,
$C$ represents the direct Coulomb interaction, and $\mathit{EX}$
represents the exchange interaction.

Defining
\begin{equation*}
S_\mathit{c,v}(\vec{r})=\sum_{b\in\{s,x,y,z\}\otimes\{\uparrow,\downarrow\}}
\chi_{c,b}^*(\vec{r})\chi_{v,b}(\vec{r})
\end{equation*}
and vector $\vec{T}$ with the components
\begin{equation*}
T^x_\mathit{c,v}(\vec{r})=\frac{P}{E_g}\sum_{\sigma\in{\uparrow,\downarrow}}
\left [ \chi_{c,s\sigma}^*(\vec{r})\chi_{v,x\sigma}(\vec{r})+
\chi_{c,x\sigma}^*(\vec{r})\chi_{v,s\sigma}(\vec{r}) \right ]
\end{equation*}
and $T^y$, $T^z$ defined analogously ($E_g$ is the fundamental
band gap and $P$ is one of the Kane's parameter related to the non-vanishing
coordinate matrix elements $\langle x | x | s \rangle=P/E_g$), we can write
\del{(see Supplementary material for a detailed derivation)}
\begin{multline*}
C(c1,v1;c2,v2)=-\frac{e^2}{4\pi\varepsilon}\times\\
\times\int d\vec{r}_1\int d\vec{r}_2
\frac{1}{|\vec{r}_1-\vec{r}_2|}S_{c1,c2}(\vec{r}_1)
S_{v2,v1}(\vec{r}_2)
\end{multline*}
($e$ denotes the elementary charge and $\varepsilon$ the dielectric function).
The exchange Coulomb interaction term $\mathit{EX}$ can be expressed as a sum
of the following three terms:
\begin{multline}
\label{ex0}
\mathit{EX}_0(c1,v1;c2,v2)=\frac{e^2}{4\pi\varepsilon}\times\\
\times \int d\vec{r}_1\int d\vec{r}_2
\frac{1}{|\vec{r}_{12}|}S_{c1,v1}(\vec{r}_1)
S_{v2,c2}(\vec{r}_2),
\end{multline}
\begin{multline}
\label{ex1}
\mathit{EX}_1(c1,v1;c2,v2)=\frac{e^2}{4\pi\varepsilon}\int d\vec{r}_1\int d\vec{r}_2
\frac{1}{|\vec{r}_{12}|^3} \times \\
\times \vec{r}_{12}\cdot
\left [ 
S_{v2,c2}(\vec{r}_2) \vec{T}_{c1,v1}(\vec{r}_1)
-S_{c1,v1}(\vec{r}_1) \vec{T}_{v2,c2}(\vec{r}_2)
 \right ],
\end{multline}
and
\begin{multline}
\label{ex2}
\mathit{EX}_2(c1,v1;c2,v2)=\frac{e^2}{4\pi\varepsilon} \int d\vec{r}_1\int d\vec{r}_2
\frac{1}{|\vec{r}_{12}|^5}\times \\ 
\times \mathop{\sum\sum}_{\alpha,\beta \in {x,y,z}}  
 T^{(\alpha)}_{c1,v1}(\vec{r}_1) T^{(\beta)}_{v2,c2}(\vec{r}_2) 
 [\delta_\mathit{\alpha\beta}|\vec{r}_{12}|^2-3{r}_{12}^{\alpha} {r}_{12}^{\beta}]
,
\end{multline}
where $\vec{r}_{12}=\vec{r}_1-\vec{r}_2$ and $\delta_\mathit{\alpha\beta}$
is the Kronecker delta.
We note that $S$ is non-zero only when the mixing of valence and conduction bands is taken into
account. Thus, only the third term of the multipole expansion, Eq.~\ref{ex2}, contributes to the FSS when this
mixing is neglected, e.g. when single band or six-band~\cite{PhysRev.97.869} k.p
theory is used to obtain the wave function of individual particles. However, as
the scaling of the terms with the linear extension of the wave function $L$
goes as $\mathit{EX}_0 \sim 1/L$, $\mathit{EX}_1 \sim 1/L^2$, $\mathit{EX}_2 \sim 1/L^3$,
the low-order terms are important in particular in larger QDs.

\section{Model of Gaussian wave functions}
\label{s3}

Before treating realistic quantum dots with the full-complexity model,
it is worth to provide an intuitive understanding of the relation between
the topology of the excitonic wave function and the value of FSS.
To this end we employed a simplified model with the
exciton composed of a single Slater determinant, neglected
band mixing, and the electron and hole densities having the form
of three-dimensional Gaussian functions. The
electron and hole envelope functions read
$$
\chi_{e,h}(\vec{r})\propto\exp\left[-\frac{(x-x_0)^2}{L_x^2}-
\frac{(y-y_0)^2}{L_y^2}-\frac{(z-z_0)^2}{L_z^2}\right]^{1/2},
$$
where $L_{x,y,z}$ determine the spatial extensions of the wave functions and
$x_0$, $y_0$, $z_0$ are the coordinates of the particle barycenter.
As the band mixing is neglected, only the dipole-dipole exchange term (Eq.~\ref{ex2})
contributes to the total FSS.

\begin{figure}
  \begin{center}
  \includegraphics[width=0.95\columnwidth]{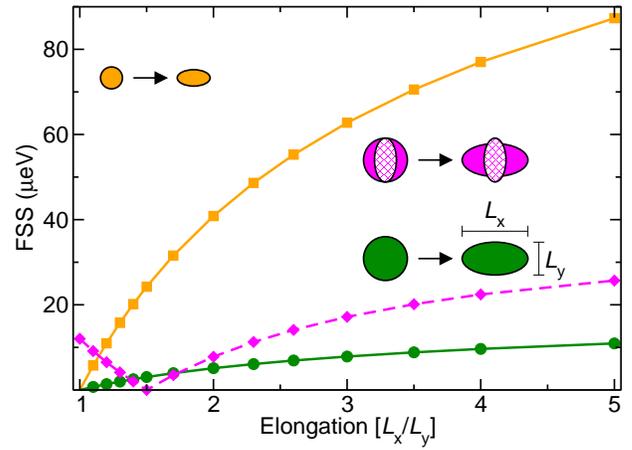}
\caption{\label{fig_model_elongation} (color online) FSS as a function of the lateral
elongation of the wave functions, defined as $E_w=L_x/L_y$. (orange line)
A smaller dot with the extension parameters
$L_x \times L_y=25\ \mathrm{nm^2}$, $L_z=2$~nm. The lateral extensions are
varied preserving the value of their product (and thus the volume 
and vertical aspect ratio of a QD).
Electron and hole wave functions are elongated equally.
(green line) A larger dot with the extensions twice larger than for the
smaller dot, i.e., $L_x \times L_y=100\ \mathrm{nm^2}$, $L_z=4$~nm.
(magenta line) A smaller dot, $L_x \times L_y=25\ \mathrm{nm^2}$, $L_z=2$~nm,
with opposite electron and hole elongation. For electrons, $E_w$ has
a constant value of 2/3. For holes, $E_w$ is varied and FSS is displayed
as a function of the hole elongation. (insets) The insets schematically
depict the topology of the wave functions. Hatched magenta ellipses
correspond to the electron wave functions.
}
  \end{center}
\end{figure}

A crucial parameter for FSS is the lateral elongation of the envelope functions defined
as $E_w=L_x/L_y$.
It follows directly from Eq.~\ref{ex2} that for non-elongated envelope functions ($L_x=L_y$)
FSS acquires a zero value. The dependence of the FSS on the elongation is shown
in Figure~\ref{fig_model_elongation}. In order to isolate the effect of the elongation
and avoid unintentional variation of other parameters, we preserved the volume
and the effective vertical aspect ratio of the model dots, i.e, values of $L_z$
and the product $L_x \times L_y$ were kept constant. First, we assumed the same
envelope function for both electron and holes (orange and green line).
Such case corresponds e.g.~to strain-free GaAs/AlGaAs dots.~\cite{PhysRevB.80.085309}
FSS exhibits a monotonically increasing concave dependence on the
lateral elongation $E_w$. To demonstrate the effect of the QD volume, we show
FSS for a smaller dot (extension parameters $L_x \times L_y=25\ \mathrm{nm^2}$, $L_z=2$~nm)
and a larger dot with two-times larger dimensions (e.g., eight-times
larger volume).
The values of FSS for a larger QD are exactly eight-times smaller. The
inverse proportionality of FSS to the QD volume or to the third power
of a characteristic linear dimension $L$ can be directly inferred
from Eq.~\ref{ex2}. With the band mixing taken into account, additional terms
proportional to $1/L$ and $1/L^2$ emerge. However, the $1/L^3$ or $1/V$ scaling
law ($V$ representing a volume of the QD) has been recently demonstrated
experimentally in realistic strain-free
GaAs/AlGaAs QDs.~\cite{PhysRevB.90.041304} FSS values exceed the natural
linewidth of the exciton recombination lines (up to units of $\mathrm{\mu eV}$)
even for a modest elongation. For example, for $E_w=1.2$ we predict FSS of
$11\ \mathrm{\mu eV}$ ($1.4\ \mathrm{\mu eV}$) in the smaller (larger) QD.
We note that for the QDs studied in Ref.~\onlinecite{PhysRevB.90.041304}
we found the values of $L_x \times L_y$ between 11 and 36~$\mathrm{nm^2}$
and $L_z$ between 1 and 2~nm.
The smaller dot case thus corresponds well to realistic GaAs/AlGaAS
QDs.

Next, we introduce an important concept of the compensated elongation.
For a suitable exciton wave function topology, FSS can attain the zero value
even in the system that lacks the required symmetry $\mathrm{C_{4v}}$.
We consider the electron envelope function to be elongated in the
direction perpendicular to the elongation of the hole envelope function,
$E_\mathit{we}=2/3$ (the subscripts $e$, $h$ are used,
when required, to distinguish the parameters of electrons and holes,
respectively). The extension parameters correspond to the smaller dot:
$L_x \times L_y=25\ \mathrm{nm^2}$, $L_z=2$~nm. 
FSS is plot as a function of the hole elongation
$E_\mathit{wh}\geq 1$ in Fig.~\ref{fig_model_elongation}
(magenta line). The prominent feature of the dependence is the zero-value
minimum at $E_\mathit{wh}=3/2$, (i.e., the inverse
of the hole elongation. Intuitively, this can be described as the mutual
compensation of both electron and hole elongations. The integral in
Eq.~\ref{ex2} attains a zero value, which is however not related to
the symmetry. In realistic QDs, the condition of the inverse elongation
does not hold (due to band mixing or different volume of the envelope
function of electrons and holes) but the effect is preserved. The minimum value
of FSS can be larger than zero in case of QDs with the irregular shape. The effect of
the compensated elongation has been already demonstrated experimentally
utilizing the anisotropic external strain to vary the elongation of the
hole envelope function.~\cite{PhysRevB.83.121302}
In the following we will demonstrate that the elongation of the
hole envelope function can be efficiently varied in type-II InAs QDs
with GaAsSb capping layer.


\begin{figure}
  \begin{center}
  \includegraphics[width=0.95\columnwidth]{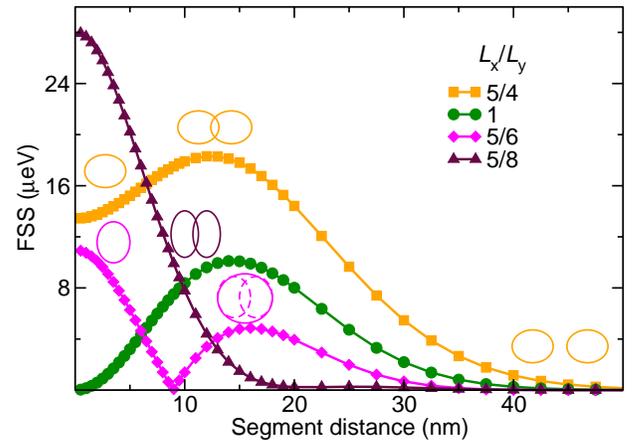}
\caption{\label{fig_model_segments} (color online) FSS for a hole wave
function composed of two segments. The extensions of both electron
and hole wave functions fulfill $L_x \times L_y=25\ \mathrm{nm^2}$, $L_z=2$~nm,
with the lateral elongation $E_w$ of 5/4 (orange line, squares),
1 (green line, circles), 5/6 (magenta line, diamonds), and 5/8 (maroon line,
triangles). The electron wave function is composed of a single Gaussian
while the hole wave function is gradually split along $x$ into two
identical segments evenly positioned around the central electron wave function.
FSS is plotted as a function of the distance between the barycenters
of the segments. Insets schematically depict the position and shape
of the segments.
}
  \end{center}
\end{figure}

The transition between type-I and type-II confinement in \gaassb{} capped
QDs is accompanied by the splitting of the hole wave function into two segments,
evenly spread along the central electron wave function in QDs with a sufficient
symmetry (Fig.~\ref{fss_lens}b,c). The behavior of FSS under
such transition is shown in Fig.~\ref{fig_model_segments}. When the connecting
line of the segments is parallel to the elongation axis (orange line),
small segment shifts from the central position effectively enhance the hole elongation and consequently
the FSS. For larger shifts the wave function disintegrates; now
the effect of increased distance of electron and hole prevails resulting
into a decrease of FSS. The same behavior is predicted for non-elongated
wave functions (green line), where the FSS dependence starts at a zero value,
increases as the holes become effectively elongated, and decreases when the
separation effects prevail.
When the connecting line of the segments and the elongation axis are
perpendicular, the effective elongation of the holes decreases as the
segments are separated, and so happens with FSS. Depending on the magnitude of
the original elongation, two possibilities exist: (1) The hole eventually
becomes elongated in opposite direction (magenta line; note the magenta insets
of Fig.~\ref{fig_model_segments} schematically depicting the change in the
elongation direction). FSS goes through a zero value and starts to increase
again. Finally, the separation effects prevail and FSS decreases. (2) When
the original elongation is large, the hole disintegrates before the elongation
direction is changed (maroon line). In such case a monotonously
decreasing dependence of FSS on the segment distance is observed, governed
first by the decrease of the effective elongation and then by the separation
effects. 

In strongly asymmetric QDs the minima in confinement potential corresponding
to both segments can differ considerably and a single-segment hole wave function
displaced from the central electron wave function can be formed. In such case
there is no effective change in the elongation and the separation effect
leads to a monotonous decrease of FSS as in the case of large perpendicular
elongation (not shown).

\section{Realistic quantum dots}
\label{s4}

We will now focus on realistic InAs QDs with a \gaassb{} capping layer.
We will show that FSS in such structures can be tuned by the thickness
of the \gaassb{} layer, allowing its decrease below the natural linewidth
of the exciton transitions.

To show the universality of the tuning approach, three QD geometries will
be considered here, denoted as pyramidal, symmetric lens-shaped, and elongated
lens-shaped.
The {\it pyramidal} QD is adopted from 
Ref.~\onlinecite{:/content/aip/journal/apl/90/21/10.1063/1.2741608} and has
the shape of a pyramid with the base length of 22~nm, height of 8~nm,
and the trumpet indium composition profile within the pyramid. For
the other structures we assume QDs composed of pure InAs. The
{\it symmetric lens-shaped} QD is modeled as a top of a sphere with the base radius of 8~nm
and the height of 4~nm. The prominent cause of the lateral asymmetry and
contributor to FSS of many
QD systems is a structural elongation. So far no elongation was reported
for InAs QDs with \gaassb{} overlayer, which is in striking contrast
with InAs QDs capped by pure GaAs.~\cite{:/content/aip/journal/apl/89/15/10.1063/1.2358845,Kuldova2006983}  
This can be attributed to the surfacting
effect of antimony but it is also possible that the elongation has been
overlooked as the methods involved in experimental studies were insensitive
to it. Therefore, we consider in our study also the possibility that QDs
are elongated. In accordance with GaAs capped InAs QDs we select the direction
[1$\bar{1}$0] as the main elongation axis and quantify the elongation
by the ratio of characteristic lateral dimensions along [1$\bar{1}$0]
and [110], denoted as $E_s$ in the following (the subscript $s$ is used to
differentiate the structural elongation from the wave-function elongation
used in the previous section). The {\it elongated lens-shaped} QDs
are formed from the lens-shaped dot by its stretching
along [1$\bar{1}$0] and compressing along [110] by the same factor so that the
QD volume and height are preserved. All QDs are capped with
the $\mathrm{GaAs_{0.8}Sb_{0.2}}$ layer of a certain thickness
and further embedded in GaAs.

\begin{figure}
  \begin{center}
    \includegraphics[width=0.95\columnwidth]{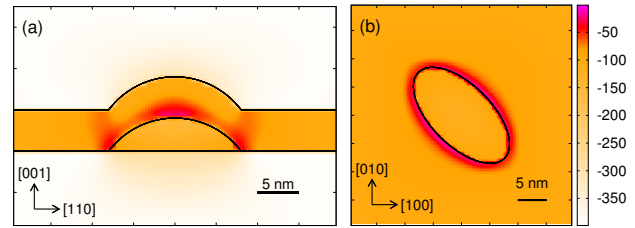}
  \end{center}
\caption{\label{potential_nop} (color online) Effective confinement potential
for holes (colorbar scale in meV) without the contribution of the piezoelectric field
for the lens-shaped QD. (a) $E_s=1$, plane ($1\bar{1}0$) through the QD center,
(b) $E_s=2$, plane (001) just above the QD base.
Boundaries between different materials are schematically depicted.
The potential is given from the electron view; the holes are confined
near the largest values represented by the red/orange spots.
}
\end{figure}

The topology of the wave functions is closely connected with the effective
confinement potential, which is contributed by the band-edge offsets,
strain field, and piezoelectric potential. We construct the potential
from the eigenvalues of the pointwise diagonalized Hamiltonian
(terms containing the spatial derivatives are discarded and
the Hamiltonian is then diagonalized at each point of the simulation grid)
so the strain-induced band mixing is already involved.
The hole potentials in all
type-II QDs discussed further in the paper exhibit qualitatively similar
features. We will present them for the exemplar lens-shaped QD with
the GaAsSb layer thickness of 5 nm and with the elongation $E_s$ of
either 1 or 2.
Figure~\ref{potential_nop} shows the potential profile without the piezoelectric
contribution. The potential is given from the electron view, the largest values
correspond to the minima of the hole confinement. Zero energy is set
to the valence band edge of bulk unstrained InAs. Two local minima 
of the hole confinement potential are formed
in the GaAsSb layer along the sides of the QD and above its top
[Fig.~\ref{potential_nop}(a)]. The top-minimum is of the
light-hole character and therefore penalized by the quantum confinement.
The ground hole state will be localized in the side-minimum which forms a ring
around the QD [Fig.~\ref{potential_nop}(b)], in
which weak variations of the potential are present with two rather shallow
absolute minima along the long QD side (i.e., in [110] direction from
the QD center). Depending on the magnitude of these variations,
the wave function might form a ring or be split into two segments.


\begin{figure}
  \begin{center}
    \includegraphics[width=0.95\columnwidth]{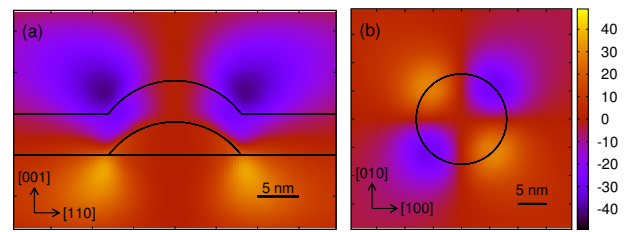}
  \end{center}
\caption{\label{potential_piezo} (color online) Piezoelectric potential
(colorbar scale in meV)
for the symmetric lens-shaped QD. (a) plane ($1\bar{1}0$) through the QD center,
(b) plane (001) just above the top of the QD (i.e., above the horizontal nodal
plane).}
\end{figure}

The piezoelectric field has an octopole shape shown in
Figure~\ref{potential_piezo}. Its contribution is rather important
as its magnitude of about 50~meV is comparable
to the variations of the rest of the confinement potential inside
the \gaassb layer.
The horizontal nodal plane of the piezoelectric octopole lies close to
the side-minimum of the confinement potential. The piezoelectric potential
therefore tends to split the wave function of the holes in the side-minimum
into two segments situated along [110] below the nodal plane
or along [$1\bar{1}0$] above the nodal plane.

\begin{figure}
  \begin{center}
    \includegraphics[width=0.95\columnwidth]{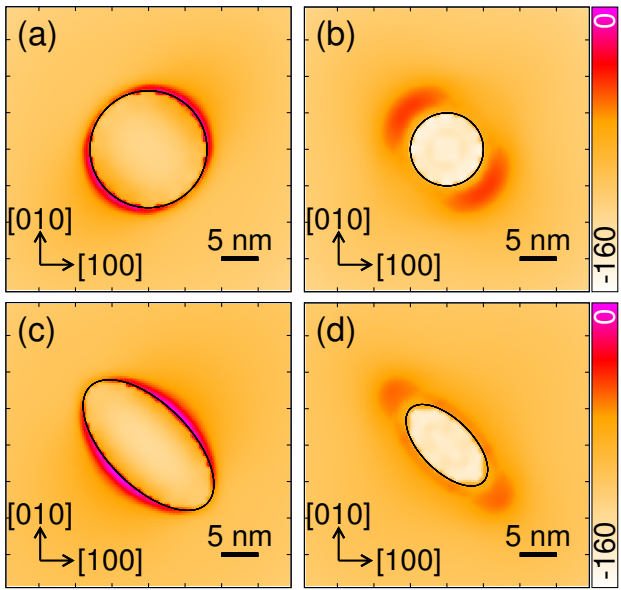}
  \end{center}
\caption{\label{potential_p1} (color online) Total confinement potential
for holes (colorbar scale in meV) including the contribution of the piezoelectric field
for the lens-shaped QD. (a,c) Plane (001) below the nodal plane of the piezoelectric potential (just above the QD base), $E_s=1$ (a) and 2 (c).
(b,d) Plane (001) above the nodal plane of the piezoelectric potential (3~nm above the QD base), $E_s=1$ (b) and 2 (d). Black circles and ellipses display the
QD boundary at respective heights.
The potential is given from the electron view; the holes are confined
near the largest values represented by the red/orange spots.}
\end{figure}

The total confinement potential is shown in Figure~\ref{potential_p1}.
The following morphologies of the hole wave function are possible: (i)
inside a QD (type I), (ii) a ring-like shape along the QD when the piezoelectric
field is too weak to localize the holes in its minima,
(iii) two segments at the dot base situated along [110] [Fig.~\ref{potential_p1}(a,c)], (iv)
two segments at the QD sides above the piezoelectric nodal plane situated along [$1\bar{1}0$] due to the
piezoelectric field [Fig.~\ref{potential_p1}(b)] or along [110] when the structural elongation along
[$1\bar{1}0$] prevails [Fig.~\ref{potential_p1}(d) is close to that case].
In pyramidal QDs with the trumpet In composition profile, both the side-minimum
of the confinement potential
and the piezoelectric octopole are shifted up towards the region of
large In content.
In short, there is a rich variety of the hole wave function morphologies.
Variation of the parameters such as the thickness or the composition
of the GaAsSb layer are supposed to induce transitions between those
morphologies. For example, switching between the deep narrow minima
below the nodal plane [Fig.~\ref{potential_p1}(a,c)] and shallow broad
minima above the nodal plane [Fig.~\ref{potential_p1}(b,d)] shall
be achievable.
Considering the effect of the compensated elongation,
this opens an interesting prospect for the tuning of FSS.

\begin{figure}
  \begin{center}
    \includegraphics[width=0.95\columnwidth]{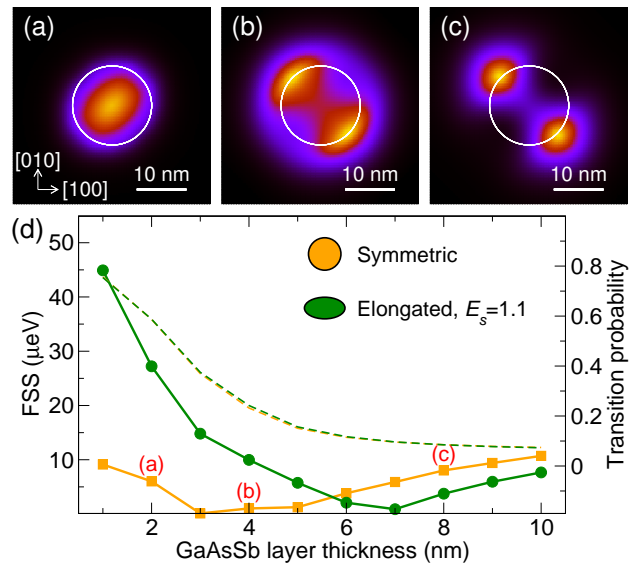}
  \end{center}
\caption{\label{fss_lens} (color online) (a,b,c) Planar probability density of the hole ground state
in the lens-shaped QD (height 4~nm, radius 8~nm, symmetric), plane (001) (integrated over [001]) for the GaAs$_{0.8}$Sb$_{0.2}$ layer
thickness of (a) 2~nm, (b) 4~nm, and (c) 8~cm. The QD boundary at the base-height is displayed with
the white line. (d) FSS (thick solid lines with symbols, left) and transition probability (thin dashed lines)
between the ground electron and hole state in the symmetric lens-shaped QD (orange) and the elongated lens-shaped
QD with $E_s=1.1$ (green) as functions of the GaAsSb layer thickness. The red labels denote the points
corresponding to the maps of the probability density.}
\end{figure}

Figure~\ref{fss_lens}(d) shows the dependence of FSS on the thickness of
the \gaassb{} layer in two lens-shaped QDs: symmetric and weakly elongated
($E_s=1.1$). We will first discuss the case of the symmetric QD. 
The electron wave function is weakly elongated in $[1\bar{1}0]$ direction
and as it resides within the QD, its variation with the thickness
of the \gaassb{} layer is negligible.
For a thin \gaassb{} layer (up to 3 nm) the ground hole wave function
resides inside the QD, too. It experiences the bottom part of the piezoelectric
octopole and is thus elongated in [110], as shown in Fig.~\ref{fss_lens}(a).
The polarization of the lower exciton component is [110].
With increasing thickness a hole ground state gradually shifts
into the \gaassb{} layer and also slightly upwards (for about 1.3~nm for the
full range of thicknesses). Consequently, it becomes
split by the upper part of the piezoelectric octopole
into two segments along $[1\bar{1}0]$. For the thickness interval
between 3 and 5~nm, the segmented wave function behaves as effectively
elongated in $[1\bar{1}0]$ [Fig.~\ref{fss_lens}(b)]
and the lower exciton component is also
polarized along $[1\bar{1}0]$. For the thickness values above 5~nm,
the segments are well separated and the elongation of each segment in [110]
determines the overall symmetry of the wave function. The lower
exciton component is again polarized along [110]. Thus, we distinguish
three regions of different exciton polarization. At each of the two transitions
between those regions, exciton levels cross and FSS is reduced to zero.
We note that minimum values obtained in our calculations are non-zero
due to the finite step in the thickness dependence and read
$0.1~\mathrm{\mu eV}$ and $1~\mathrm{\mu eV}$ for the first and second
transition, respectively.
The observed behavior of FSS corresponds well to the qualitative
prediction of the model of Gaussian function for $L_x/L_y=5/6$
(cmp. Fig.~\ref{fig_model_segments}, magenta line).

\begin{figure}
  \begin{center}
    \includegraphics[clip,width=0.95\columnwidth]{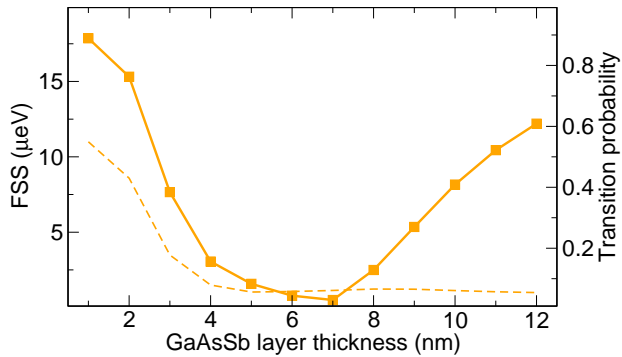}
  \end{center}
\caption{\label{fss_pyramid} (color online) FSS (thick solid line with symbols, left) and transition probability (thin dashed line)
between the ground electron and hole state in the pyramid-shaped QD as functions of the GaAsSb layer thickness.}
\end{figure}

Similar behavior is observed in a lens-shaped QD weakly elongated in
$[1\bar{1}0]$ [Figure~\ref{fss_lens}(d), green lines].
The first region is now missing, as the elongation of the hole
wave function within the QD volume (for a thin \gaassb{} layer) is now
$[1\bar{1}0]$ because the structural elongation dominates over
the piezoelectric contribution. The zero FSS is reached for a \gaassb{}
layer thickness of 7 nm. Figure~\ref{fss_pyramid} shows the FSS
dependence on the thickness of the SRL layer in a pyramidal QD with the
trumpet profile (height of 8 nm and base size of 22 nm). As in the case
of the lens-shaped QDs, the hole function changes its effective elongation
during the crossover between the type-I and type II confinement and in turn,
exciton levels cross and FSS is reduced towards zero. Thus, variations
of the \gaassb{} layer thickness present a universal approach to tune and
reduce FSS. Further, its tuning power is considerable as it can compensate
even a structural elongation, as was demonstrated for the elongated
lens-shaped QD.
For a typical lifetime of InAs QDs of 1~ns, the natural
linewidth of the exciton recombination is $4\ \mathrm{\mu eV}$;
experimentally predicted spectral linewidth approaches 
$100\ \mathrm{\mu eV}$.~\cite{0957-4484-22-6-065302} 
Both values are larger than the minimum accessible values predicted
here for the \gaassb{} capped QDs.

\del{ 
\cc{Nasledujici odstavec mozna vypustit, pokud ho nedopocitame. Kdyztak si o to rekne referee.}
A caution has to be used when applying results obtained for idealized
QDs to realistic QDs whose shape is always somewhat distorted.
In our previous work~\cite{PhysRevB.83.121302} we have modeled such a distortion
by tilting the wave function with respect to the crystallographic axes;
the minimum accessible FSS in strongly elongated and slightly irregular QD was $4\ \mathrm{\mu eV}$.
\cc{mozna dopocitat pro naklonenou pyramidu}.
Further, wave function are in general less irregular
than the structure; smooth wave functions are enforced by the finite mass.
For a typical lifetime of InAs QDs of 1~ns, the natural
linewidth is $4\ \mathrm{\mu eV}$ which is comparable with
our estimate of the minimum accessible values in the \gaassb{} capped QDs.
}

Finally we argue that \gaassb{} capped InAs QDs are good emitters even in type-II confinement
regime, contrasting the other type-II material systems in which the studies of the
optical properties of individual QDs are rare and
challenging.~\cite{:/content/aip/journal/apl/90/1/10.1063/1.2425039,PhysRevB.86.115305}
There is an experimental evidence of intense photoluminescence from the
\gaassb{} capped InAs QDs,~\cite{:/content/aip/journal/apl/101/25/10.1063/1.4773008}
about 4-times weaker than for the reference type-I sample and about 3-times stronger
after rapid thermal annealing. Roughly 15-times higher exciton lifetime
has been reported (0.7~ns in type I compared to 11.2~ns in type II).
We support these observations with the calculated transition probabilities (square moduli of matrix elements) shown
in Figs.~\ref{fss_lens}(d) and~\ref{fss_pyramid}. The values are normalized to the full overlap
of the envelope functions. Typical values for type I QDs are in the range 0.5\,--0.7.
The values corresponding to the minima of FSS are 0.37 and 0.15 for the symmetric lens-shaped QD,
0.10 for the elongated lens-shaped QD, and 0.12 for the pyramidal QD. Thus, roughly 5-times
weaker photoluminescence as compared to type-I QDs is expected.
For this reason we are convinced that the extraction of individual photons
or photon pairs from individual \gaassb{} capped QDs
will be experimentally feasible.

\section{Conclusions}
\label{s5}
In \gaassb{} capped InAs QDs, lateral symmetry of the hole wave functions
can be to a large extent influenced by the thickness of the \gaassb{}
layer. In particular, during the crossover between type-I and type-II
confinement regimes, the hole wave function is shifted upwards
across the nodal plane of the piezoelectric octopole, which is accompanied
by the change of the direction of lateral elongation. Due to the mechanism
of compensated elongation, a crossing of the bright exciton levels and
a reduction of FSS to zero is predicted for certain thicknesses of
the \gaassb{} layer. Low natural FSS and efficient photoluminescence
make the \gaassb{} capped InAs QDs attractive as a possible
source of entangled photon pairs.

\section{Acknowledgment}
This work was supported by European Social Fund (grant
No.~CZ.1.07/2.3.00/30.0005),
European Regional Development Fund (project
No.~CZ.1.05/1.1.00/02.0068),
the Grant Agency of the Czech Republic 
(grant No. 15-21581S), Technology Agency of the Czech Republic
(grant No. TE01020233), and EU 7th Framework Programme (Contracts No. 286154
-- SYLICA and 280566 -- UnivSEM). 
P.~K.~was supported by the internal project MUNI/A/1496/2014.

\bibliography{FSSII}

\end{document}